\documentclass[aps,prl,twocolumn,groupedaddress,preprintnumbers]{revtex4-1}

\usepackage{amssymb,amsmath,hhline,graphicx}
\usepackage{diagbox}
\usepackage{autobreak}

\newcommand{\nn}{\nonumber \\}

\begin{document}

\preprint{ZU-TH 48/18}

\title{Exact Top Yukawa corrections to Higgs boson decay into bottom quarks}


\author{Amedeo Primo}

\email[]{aprimo@physik.uhz.ch}
\affiliation{Department of Physics, University of Z{\"u}rich, CH-8057 Z{\"u}rich, Switzerland}
\author{Gianmarco Sasso}

\email[]{sasso.gianmarco@gmail.com}
\affiliation{Universit\`a di Napoli Federico II,\\
Complesso di Monte Sant'Angelo, Edificio 6 \\
via Cintia I-80126, Napoli, Italy}
\author{G\'abor Somogyi}

\email[]{gabor.somogyi@cern.ch}
\affiliation{MTA-DE Particle Physics Research Group,
H-4010 Debrecen, PO Box 105, Hungary}
\author{Francesco Tramontano}

\email[]{francesco.tramontano@unina.it}
\affiliation{Universit\`a di Napoli Federico II
and INFN sezione di Napoli,\\
Complesso di Monte Sant'Angelo, Edificio 6 \\
via Cintia I-80126, Napoli, Italy}
%

\date{\today}

\begin{abstract}
In this letter we present the results of the exact computation 
of contributions to the Higgs boson decay into bottom quarks 
that are proportional to the top Yukawa coupling. Our 
computation demonstrates that approximate results already 
available in the literature turn out to be particularly 
accurate for the three physical mass values of the Higgs 
boson, the bottom and top quarks. Furthermore, contrary to 
expectations, the impact of these corrections on differential
distributions relevant for the searches of the Higgs boson
decaying into bottom quarks at the Large Hadron 
Collider is rather small.
\end{abstract}


\maketitle

\section{Introduction}

The discovery of the Higgs boson~\cite{Higgs:1964ia,Englert:1964et} by the ATLAS~\cite{Aad:2012tfa}
and CMS~\cite{Chatrchyan:2012xdj} experiments at CERN has 
ushered in a new era in particle physics phenomenology. 
The Standard Model (SM) of elementary particles is now 
complete and there is no decay or scattering phenomenon at low energies that significantly deviates from what is predicted by the SM. Still, we know that the SM cannot be the ultimate theory, if not for the lack of consistency with mainly cosmological observations, but for the fact that it contains quite a large number of parameters, which makes it unreasonable 
to think of it as a fundamental theory.
The LHC is guiding the experimental community towards the study of the 
Higgs potential and the Higgs boson direct couplings with all the other particles, an experimentally previously completely 
unexplored sector of the SM Lagrangian.
The SM makes very precise predictions for all the vertices and couplings of the Higgs boson and the verification of these predictions is among the fundamental questions addressed at 
the LHC.

In particular, the gluon fusion production mechanism has 
given direct access to Higgs boson decay into vector bosons 
and an indirect access the top Yukawa coupling. More recently, also the direct coupling of the Higgs boson to the top quark has been observed~\cite{Aaboud:2018urx,Sirunyan:2018shy}.
Measurements of the Higgs coupling to the tau lepton have been extracted by combining all production modes~\cite{Sirunyan:2017khh,Aaboud:2018pen}, while the direct coupling to the bottom quark
has been observed by exploiting the features of the VH (V
= W or Z) associated
production mechanism~\cite{Aaboud:2018zhk,Sirunyan:2018kst}. The decay to bottom quarks is quite special, because
it is the one with the largest branching ratio.
The decay width of the Higgs boson into bottom quarks has been
computed at up to four loops in QCD~\cite{Gorishnii:1990zu,Gorishnii:1991zr,Kataev:1993be,Surguladze:1994gc,Larin:1995sq,Chetyrkin:1995pd,Chetyrkin:1996sr,Baikov:2005rw} using an approximated treatment of the 
bottom quark mass, up to one loop including electroweak corrections~\cite{Dabelstein:1991ky,Kniehl:1991ze} and also including mixed QCD-electroweak effects~\cite{Mihaila:2015lwa}. The exact bottom mass corrections have been computed up to two loops~\cite{Bernreuther:2018ynm}.
A relatively large component of the two loop computation is represented by the diagrams in which the Higgs boson couples 
to a top quark loop, see Figs.~\ref{fig:diag2L} and~\ref{fig:real}. These two sets of diagrams are both UV and IR finite separately, and their contributions have been computed only approximately in~\cite{Chetyrkin:1995pd}, finding a very compact formula
that should be considered valid for values of the masses such that $m_b \ll m_H \ll m_t$. Comparing this formula with the 
rest of the two loop contributions, it turns out that these 
pieces, proportional to the top Yukawa coupling $y_t$, account for about $30\%$ of 
the total two loop result.

The aim of this letter is twofold. First, we want to assess 
the impact of the neglected terms in the expansion of~\cite{Chetyrkin:1995pd}, that in principle could be of 
the order of $(m_H/m_t)^4\sim 20\%$ (see Eq.~(3) in~\cite{Chetyrkin:1995pd}). We do this by computing the 
full analytic result for the contributions to the Higgs boson 
decay into bottom quarks that are proportional to the top 
Yukawa coupling, including the exact dependence on the top 
and bottom quark masses. Furthermore, recently two groups have computed the fully differential decay width of the Higgs boson into
bottom quarks up to two loops for the massless case, and merged this computation to the two loop corrections to
the associated production in hadronic collisions~\cite{Ferrera:2017zex,Caola:2017xuq}. The corrections 
to key distributions like the transverse momentum 
and the mass spectra of the Higgs boson (reconstructed using 
the two hardest b-jets in the final state) are found to be very large.
$y_t$ contributions to the Higgs decay into bottom quarks are 
not included in the differential results mentioned above and so it
is natual to ask what the
impact of these corrections
is (see for example~\cite{Caola:2017xuq}). 
We answer this second question by presenting differential 
results which include the $y_t$ contributions to the decay 
and retain the full mass dependence on the top and bottom 
quark masses.


\section{Calculation}
\subsection{Double virtual}
 \begin{figure}
 \includegraphics[width=0.4\textwidth]{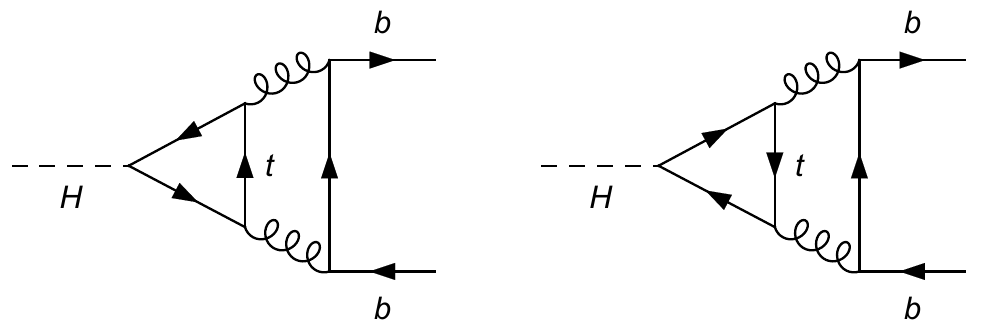}%
 \caption{Virtual $\mathcal{O}(\alpha_s^2y_t)$ contributions 
 to $H\to b\bar b$ decay.\label{fig:diag2L}}
 \end{figure}
The decay $H(q)\to b(p_1) + \bar b(p_2)$ receives $\mathcal{O}(\alpha_s^2y_t)$ contributions from the interference $|\mathcal{M}_{y_t,b\bar{b}}|^2$ of the Born amplitude
with two loop virtual corrections that involve a closed top quark loop,
\begin{equation}
|\mathcal{M}_{y_t,b\bar{b}}|^2\equiv2\text{Re}\,\mathcal{M}^{(0)\,\dagger}_{b\bar{b}}\mathcal{M}^{(2)}_{y_t,b\bar{b}}\,,
\label{eq:interf}
\end{equation}
where $\mathcal{M}^{(2)}_{y_t,b\bar{b}}$ is given by the two diagrams shown in Fig.~\ref{fig:diag2L}. 
By evaluating the Feynman diagrams, we decomposed 
$|\mathcal{M}_{y_t,b\bar{b}}|^2$ as
\begin{equation}
|\mathcal{M}_{y_t,b\bar{b}}|^2=\alpha_s^2y_tC_{A}C_{F}\sum_{\vec{a}} c_{a_1\cdots a_7}(\epsilon, m_i^2)\,I_{a_1\cdots a_7}(\epsilon, m_i^2)\,,
\label{eq:dec}
\end{equation}
with $a_j\in \mathbb{Z}$ and $i=t,b,H$.
In Eq.~\eqref{eq:dec}, the $c_{a_1\cdots a_7}$ are rational coefficients and $I_{a_1\cdots a_7}$ are two loop integrals of the type
\begin{equation}
I_{a_1\cdots a_7}(\epsilon,\!m_i^2)\!=\!\int\! \frac{\mathrm{d}^d k_1}{(2\pi)^d}\!\frac{\mathrm{d}^d k_2}{(2\pi)^d}\,
 \! \frac{D_7^{a_7}}{D_{1}^{a_1}D_{2}^{a_2}D_{3}^{a_3}D_{4}^{a_4} D_{5}^{a_5}D_{6}^{a_6}}\,,
  \label{eq:family}
\end{equation}
defined by the set of inverse propagators:
\begin{gather}
D_1 = k_1^2-m_t^2,\quad
D_2 = k_2^2,\quad
D_3 = (k_1-k_2)^2-m_t^2,\nn
D_4 = (k_1+q)^2-m_t^2, \quad
D_5 = (k_2+q)^2,\nn
D_6 = (k_2+q-p_1)^2-m_b^2\quad
D_7 = (k_1+q-p_1)^2
\label{eq:Ds}
\, ,
\end{gather}
with kinematics $m_H^2=q^2=(p_1+p_2)^2$, $p_1^2=p_2^2=m_b^2$.

We computed the loop integrals through the consolidated differential equations (DEs) method~\cite{Kotikov:1990kg,Remiddi:1997ny,Gehrmann:1999as,Argeri:2007up}.
First, we used integration-by-parts identities (IBPs)~\cite{Tkachov:1981wb,Chetyrkin:1981qh,Laporta:2001dd}, generated with the help of \textsc{Reduze2}~\cite{vonManteuffel:2012np}, in order to reduce the integrals that appear in $|\mathcal{M}_{y_t,b\bar{b}}|^2$ to a set of 20 independent master integrals (MIs) $\vec{\mathcal{I}}=(\mathcal{I}_I\,,\dots\,,\mathcal{I}_{20})$.
Subsequently, we derived the analytic expression of the MIs by solving the system of coupled first-order DEs in the kinematic ratios $m_H^2/m_t^2$ and $m_b^2/m_t^2$.
The structure of the DEs, and hence of their solutions, is simplified by parametrizing such ratios in terms of the variables $x$ and $y$, defined by
\begin{equation}
\frac{m_H^2}{m_t^2}=-\frac{(1-x^2)^2}{x^2}\,,\quad \frac{m_b^2}{m_t^2}=\frac{(1-x^2)^2}{(1-y)^2}\frac{y^2}{x^2}\,,
\label{eq:vars}
\end{equation}
and by using the Magnus exponential method~\cite{Argeri:2014qva,DiVita:2014pza,Bonciani:2016ypc,DiVita:2017xlr,Mastrolia:2017pfy,DiVita:2018nnh} in order to identify a basis of MIs that fulfil a system of canonical DEs~\cite{Henn:2013pwa},
\begin{equation}
\mathrm{d}\, \vec{\mathcal{I}}=\epsilon\,\mathrm{d}\mathbb{A}\,\vec{\mathcal{I}}\,,\quad\text{with}\;\;\mathrm{d}f=\sum_{z=x,y}\mathrm{d}\, z\,\frac{\partial}{\partial z}f\,.
\label{eq:canonical}
\end{equation}
In Eq.~\eqref{eq:canonical}, the coefficient matrix $\mathrm{d}\mathbb{A}$ is a $\mathrm{d}\!\log$-form that contains 12 distinct letters,
\begin{align}
\mathrm{d}\mathbb{A}&=
\mathbb{M}_1\, \mathrm{d}\!\log{(x)}
+\mathbb{M}_2\, \mathrm{d}\!\log{(1+x)}
+\mathbb{M}_3\, \mathrm{d}\!\log{(1-x)}\nn
&+\mathbb{M}_4\, \mathrm{d}\!\log{(1+x^2)}
+\mathbb{M}_5\, \mathrm{d}\!\log{(y)}
+\mathbb{M}_6\, \mathrm{d}\!\log{(1+y)}\nn
&+\mathbb{M}_7\, \mathrm{d}\!\log{(1-y)}
+\mathbb{M}_8\, \mathrm{d}\!\log{(1+y^2)}\nn
&+\mathbb{M}_9\, \mathrm{d}\!\log{(x+y)}
+\mathbb{M}_{10}\, \mathrm{d}\!\log{(x-y)}\nn
&+\mathbb{M}_{11}\, \mathrm{d}\!\log{(1+xy)}
+\mathbb{M}_{12}\, \mathrm{d}\!\log{(1-xy)}\,,
\label{eq:dlog}
\end{align}
with $\mathbb M_{i}\in \mathbf{M}^{20\times20}(\mathbb{Q})$. Since all letters are algebraically-rooted polynomials, we could derive the $\epsilon$-expansion of the general solution of Eq.~\eqref{eq:canonical} in terms of two-dimensional generalized polylogarithms (GPLs)~\cite{Goncharov:polylog,Remiddi:1999ew,Gehrmann:2001pz,Gehrmann:2001jv,Vollinga:2004sn}, by iterative integration of the $\mathrm{d}\!\log$-form, which we performed up $\mathcal{O}(\epsilon^4)$, i.e. to GPLs of weight four.
In order to fully specify the analytic expression of the MIs, we complemented the general solution of DEs with a suitable set of boundary conditions. The latter were obtained by demanding the regularity of the MIs at the pseudo-thresholds $m_{H}^2=0$ and $m_{H}^2=4m_b^2$ that appear as unphysical singularities of the DEs. 


The expression of the MIs obtained in this way is valid in the Euclidean region $m_H^2<0\,\land \,0<m_b^2<m_t^2$, where the logarithms of Eq.~\eqref{eq:dlog} have no branch-cuts and, 
hence, the MIs are real. The values of the MIs for positive values of the Higgs squared momentum, and in particular for the decay region $m_H^2>4m_b^2$, are obtained through analytic continuation, by propagating the Feynman prescription $m_H^2\to m_H^2+i0^+$ to the kinematic variables $x$ and $y$. 
All results have been numerically validated with \textsc{GiNaC}~\cite{Bauer:2000cp} against the results of \textsc{SecDec3}~\cite{Borowka:2015mxa}, both in the Euclidean and in the physical regions.

Upon inserting the expressions of the MIs into Eq.~\eqref{eq:dec}, we observed the expected analytic 
cancellation of all $\epsilon$-poles and obtained a 
finite result for $|\mathcal{M}_{y_t,b\bar{b}}|^2$,
\begin{equation}
|\mathcal{M}_{y_t,b\bar{b}}|^2=\frac{\alpha_s^2}{\pi^2}C_{A}C_{F}y_t\,y_b\,m_t\,m_b\,\text{Re}\,\mathcal C(x,y)\,,
\label{eq:intres}
\end{equation}
with $\mathcal C(x,y)$ being a polynomial combination, with algebraic coefficients, of 256 distinct GPLs. The explicit expression of $\mathcal C(x,y)$, as well as of the newly computed MIs, can be released by the authors upon request.

\subsection{Real-virtual}
 \begin{figure}
 \includegraphics[width=0.35\textwidth]{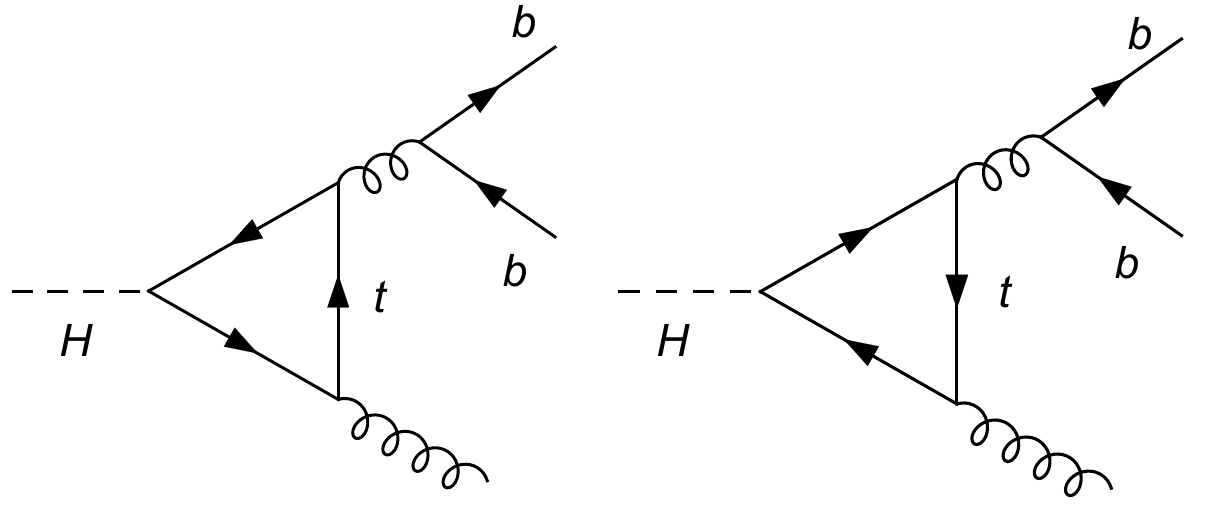}%
 \caption{Real-virtual $y_t$ contributions to $H\to b\bar b$ decay.\label{fig:real}}
 \end{figure}
The real-virtual part of the computation involves the 
interference $|\mathcal{M}_{y_t,b\bar{b}g}|^2$ of the tree 
level amplitude for $H(q) \to b(p_1) + \bar{b}(p_2) + g(p_3)$ with 
the loop diagrams in Fig.~\ref{fig:real} containing a 
closed top quark loop,
\begin{equation}
    |\mathcal{M}_{y_t,b\bar{b}g}|^2 \equiv
    2\text{Re}\,\mathcal{M}^{(0)\,\dagger}_{b\bar{b}g}
    \mathcal{M}^{(1)}_{y_t,b\bar{b}g}\,.
\label{eq:interfg}
\end{equation}
We used standard techniques to evaluate the one loop 
amplitude. As expected, $|\mathcal{M}_{y_t,b\bar{b}g}|^2$ 
is finite (in $\epsilon$) and can be written as
\begin{widetext}
\begin{align}
    |\mathcal{M}_{y_t,b\bar{b}g}|^2 &= 32 \alpha_s^2 C_{A}C_{F}y_t\,y_b\,m_t\,m_b
    \Bigg(\frac{4(s_{12}+2m_b^2)}{(s_{13}+s_{23})^2}
    + \frac{s_{13}+s_{23}-4m_b^2}{s_{13}s_{23}}\Bigg)
    \nn &\times
    \Bigg\{2\Bigg[\sqrt{\frac{4m_t^2}{m_H^2}-1}
    \Bigg(\arctan\sqrt{\frac{4m_t^2}{m_H^2}-1}
    -\frac{\pi}{2}\Bigg)
    -\sqrt{\frac{4m_t^2}{s_{12}+2m_b^2}-1}
    \Bigg(\arctan\sqrt{\frac{4m_t^2}{s_{12}+2m_b^2}-1}
    -\frac{\pi}{2}\Bigg)\Bigg]
    \nn &
    +\frac{s_{13}+s_{23}}{s_{12}+2m_b^2}
    \Bigg[1 - \Bigg(\frac{4m_t^2}{s_{13}+s_{23}}-1\Bigg)
    \Bigg[
    \Bigg(\arctan\sqrt{\frac{4m_t^2}{m_H^2}-1}
    -\frac{\pi}{2}\Bigg)^2
    -\Bigg(\arctan\sqrt{\frac{4m_t^2}{s_{12}+2m_b^2}-1}
    -\frac{\pi}{2}\Bigg)^2
    \Bigg]\Bigg]
    \Bigg\}\,,
    \label{eq:Mytbbg}
\end{align}
\end{widetext}
plus terms that vanish in four dimensions. In Eq.~\eqref{eq:Mytbbg}, 
$s_{ij}$ denotes twice the dot-product of momenta, $s_{ij} \equiv 2p_i\cdot p_j$. 


We integrated the real-virtual 
contribution over the whole phase space both analytically 
and numerically using Monte Carlo integration, finding 
perfect agreement. 
The analytic computation was performed 
by direct integration of $|\mathcal{M}_{y_t,b\bar{b}g}|^2$ 
over the three-particle phase space. 
The phase space measure for the decay 
$H(q) \to b(p_1) + \bar{b}(p_2) + g(p_3)$ reads
\begin{align}
  \mathrm{d} PS_3 &=
  2^{-10+6\epsilon} \pi^{-5+4\epsilon} (q^2)^{-1+\epsilon}
  (\Delta_3)^{-\epsilon} 
  \nn &\times
  \Theta(\Delta_3)
  \delta(q^2 - 2m_b^2 - s_{12} - s_{13} - s_{23})
  \nn &\times
  \mathrm{d}\Omega_{d-1}\mathrm{d}\Omega_{d-2}
  \mathrm{d}s_{12} \mathrm{d}s_{13} \mathrm{d}s_{23}\,,
  \label{eq:PS3}
\end{align}
where $\Delta_3$ is given by
\begin{equation}
  \Delta_3 = s_{12} s_{13} s_{23} 
  - m_b^2\left(s_{13}^2 + s_{23}^2\right)\,.
\end{equation}
The integral is finite in four dimensions and 
was evaluated in terms of GPLs after suitable 
transformations of the integration variables. 
In particular, square roots involving the integration 
variables appear at intermediate stages of the calculation 
(both from the one loop matrix element and from resolving 
the phase space constraint implied by the positivity 
of $\Delta_3$) and must be linearized, e.g. by using the 
techniques of \cite{Besier:2018jen}. The full result is 
represented in terms of a formula with 1841 distinct GPLs and 
can be released by the authors upon request. 

The numerical integration of the real-virtual contribution 
is straightforward and has been used to validate the analytic 
computation. It also allows to build Monte Carlo simulations 
with acceptance cuts and has been used to obtain the 
differential result of the next section.

\section{Results}
We begin the presentation of our results by discussing the inclusive decay rate. In Table~\ref{tab:comparison}, we compare our exact formula, obtained from the sum of the double virtual and real-virtual contributions described in the previous section, to the approximated one of Ref.~\cite{Chetyrkin:1995pd}.
The numbers in the table are obtained with the following formula for the
relative discrepancy among exact and approximated results:
\begin{equation}
    \mathbf{d}=100\,\left(1 - \frac{\Gamma_{y_t}^{Approx}}{\Gamma_{y_t}^{Exact}}\right)\,.
\end{equation}
The agreement is excellent for the physical mass values, proving for the first
time and in a completely independent way the validity of the approximated formula, and the
fact that it works much better then expected.

We now turn to the second question regarding the impact of 
the $y_t$ contribution at differential level. To this extent, 
we present results for Higgs boson associated
production and to avoid the contamination from initial state 
radiation we consider $pp\to W^+(l^+\nu_l)H(b\bar{b})$ 
at leading order and add the corrections to the decay process at the next-to-leading order. Then, we
compare this result with the one obtained by adding also the $y_t$ contribution.
Note that in both cases we normalize the cross section to the total Higgs boson decay
width into bottom quarks reported in the Yellow Report of 
the Higgs Cross Section Working Group~\cite{deFlorian:2016spz} (HXSWG), that includes higher order corrections.
So, effectively, we are comparing the shapes of distributions.
To obtain our results, we use the SM parameters recommended 
by the HXSWG and the NNPDF3.0~\cite{Ball:2014uwa} parton distribution functions. Furthermore, we impose the following typical lepton acceptance cuts: selected events must have a missing
transverse momentum greater than $30\,$GeV, the charged lepton is required to have a transverse
momentum greater than $15\,$GeV and an absolute rapidity smaller than $2.5$ and, finally, the W boson
transverse momentum is required to be larger than $150\,$GeV.
We reconstruct jets using the anti-kt algorithm with the resolution variable set to $0.5$
and require two b-jets with transverse momentum greater than $25\,$GeV and
absolute rapidity smaller than $2.5$.
\begin{table}[htb]
	\caption{The discrepancy $\mathbf{d}$ between our result and the approximate formula in \cite{Chetyrkin:1995pd}, we fix $m_b=4.92\,$GeV. 
    }
\begin{tabular}{c|c|c|c|c}
	\backslashbox{$m_t$}{$m_H$} & 20 & 75 & 125 & 180  \\ \hline
	100	&	2.123  &  0.075  &	1.025 &  6.704   \\ \hline    
	125&    2.329  &  0.011	&	0.335 &  2.107  \\ \hline	
	175	&	2.452  &  -0.019	&	0.018 &  0.355  \\ \hline	
	250	&	2.566  &  -0.024	&	-0.055 &  -0.035  \\ \hline       
	350	&	2.656  &  -0.023	&	-0.069 &  -0.113  
\end{tabular} 
\label{tab:comparison}
\end{table}

\begin{figure}[h!]
    \centering
    \includegraphics[scale=0.65,page=1]{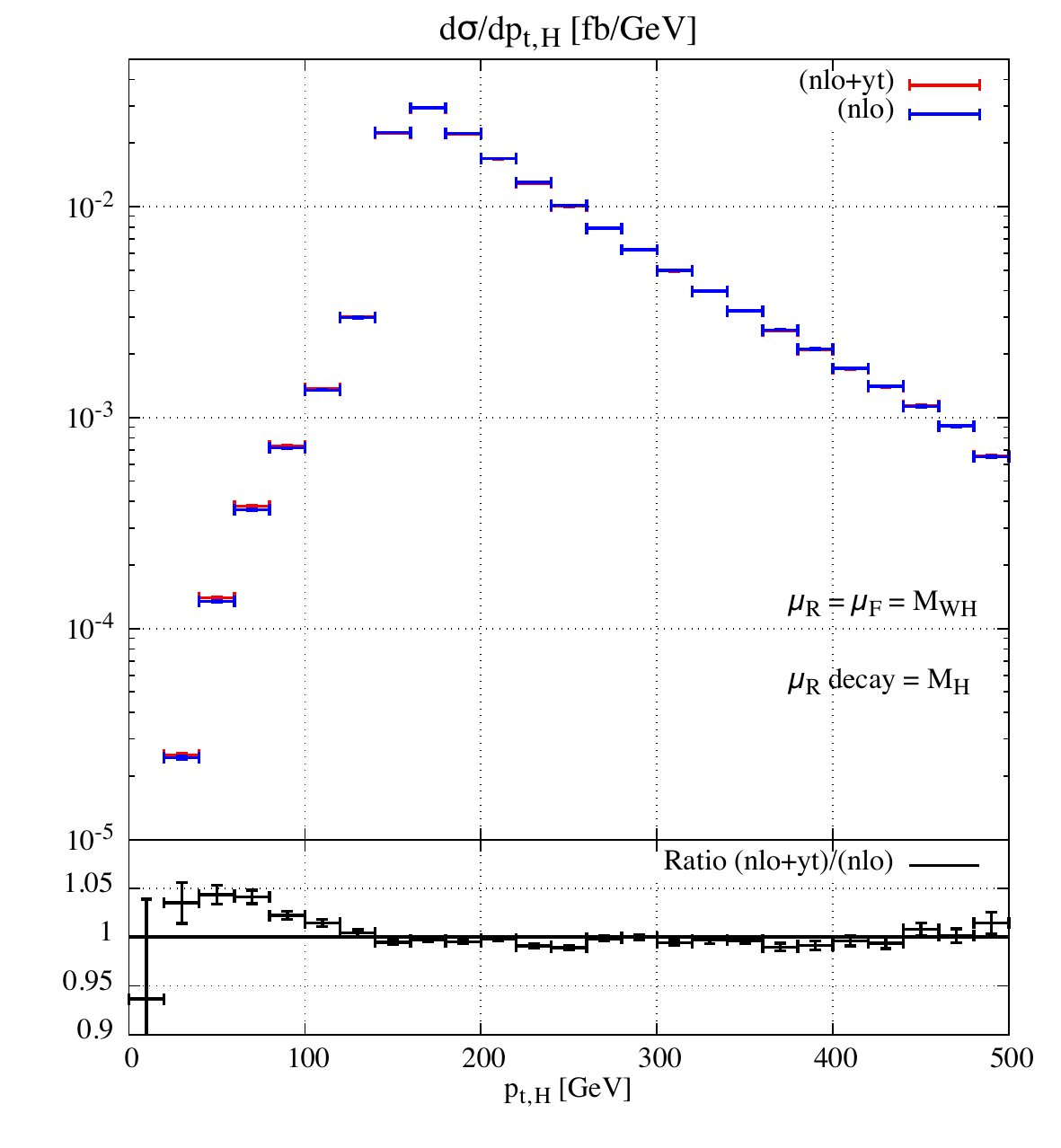}
    \caption{Transverse momentum distribution of the two b-jet system from $WH(bb)$ production in proton proton collisions at $14\,$TeV. Only corrections to the Higgs decay into bottom quarks are included.}
    \label{fig:res1}
\end{figure}
\begin{figure}[h!]
    \centering
    \includegraphics[scale=0.65,page=2]{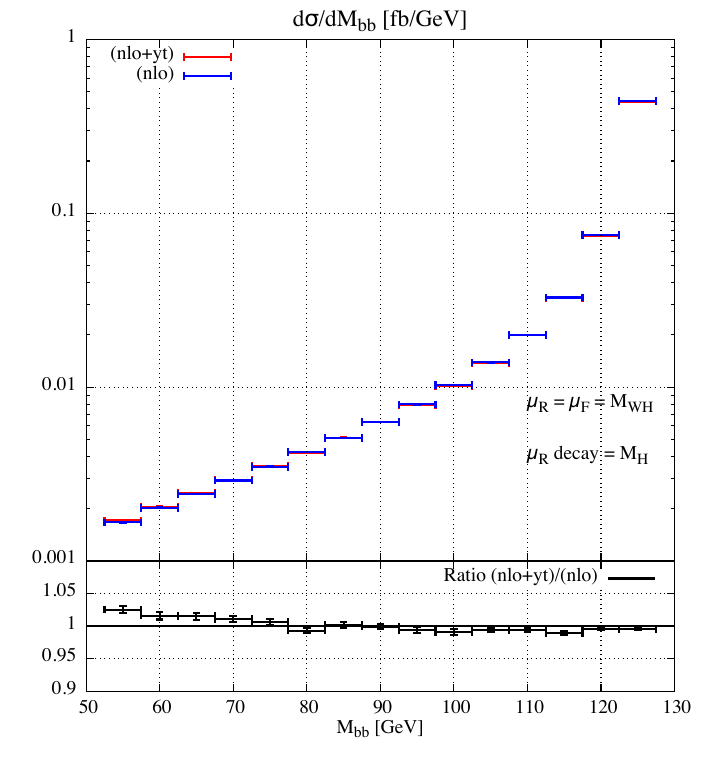}
    \caption{Mass distribution of the two b-jet system from $WH(bb)$ production in proton proton collisions at $14\,$TeV. Only corrections to the Higgs decay into bottom quarks are included.}
    \label{fig:res2}
\end{figure}
In Figs.~\ref{fig:res1} and~\ref{fig:res2} we present the 
transverse momentum and mass distributions of the two 
b-jet system from $WH(bb)$ production at the $14\,$TeV LHC. 
The error bars in the figures represent the statistical 
uncertainty associated with Monte Carlo integration.
We observe that the impact of the $y_t$ contribution on 
both the transverse momentum distribution and the mass 
distribution of the putative Higgs boson is extremely 
small with at most a $5\%$ effect in the low energy tail
of the transverse momentum distribution. These corrections 
are much smaller than the scale variation uncertainty of the computation.

\section{conclusions}
In this letter we presented the results of the full analytic computation of the top Yukawa contribution to the Higgs boson decay width into bottom quarks. First, we demonstrated that 
the approximate formula used so far in the literature works 
exceedingly well for physical values of the masses. This 
nice behaviour was not predictable a priori and, with respect 
to a possible estimate of about a $20\%$ error, we have instead found smaller than per mill deviations of this formula from the
exact result. Then, we showed that the impact of this contribution at the differential level is very small and Monte Carlo simulations performed so far are not affected by an additional significant source of uncertainty due to the neglected terms proportional to $y_t$.



%





\begin{acknowledgments}
This work has been supported by the Swiss National Science Foundation under grant number 200020-175595 (AP), the Italian Ministry of Education and Research MIUR, under project n$^{\rm o}$ 2015P5SBHT, by the INFN Iniziativa Specifica ENP (FT) and by grant K 125105 of the National Research, Development and Innovation Fund in Hungary (GS).
\end{acknowledgments}

\bibliography{references}

\end{document}